\documentclass[11pt]{article}
\usepackage{latexsym}
\usepackage{amssymb}
\usepackage{amsmath}
\usepackage{graphicx}
\usepackage{subfigure} 
\usepackage{amsfonts}

\numberwithin{equation}{section}

\newcommand{\be}{\begin{equation}}
\newcommand{\ee}{\end{equation}}
\newcommand{\ba}{\begin{eqnarray}}
\newcommand{\ea}{\end{eqnarray}}
\newcommand{\baa}{\begin{array}}
\newcommand{\eaa}{\end{array}}

\newcommand{\nr}[1]{(\ref{#1})}

\def\gsim{\raise0.3ex\hbox{$>$\kern-0.75em\raise-1.1ex\hbox{$\sim$}}}
\def\lsim{\raise0.3ex\hbox{$<$\kern-0.75em\raise-1.1ex\hbox{$\sim$}}}

\newcommand{\leff}{{\lambda_{\mathrm{eff}}}}
\newcommand{\mmuu}{{\bar{\mu}}}

\newcommand{\BbbR}{{\mathbb{R}}}
\newcommand{\BbbZ}{{\mathbb{Z}}}

\DeclareMathOperator{\sgn}{sgn}

\begin{document}
\begin{titlepage}
\renewcommand{\thefootnote}{\fnsymbol{footnote}}
\renewcommand{\baselinestretch}{1.3}
\medskip

\begin{center}
{\Large {\bf Polymer quantization, singularity resolution and\\[1ex]
the $1/r^2$ potential}}
\vspace{1cm} 


\renewcommand{\baselinestretch}{1}
{\bf
Gabor Kunstatter${}^\dagger$, 
Jorma Louko${}^\sharp$
and
Jonathan Ziprick${}^\ddagger$
\\}
\vspace*{0.7cm}
{\sl
${}^\dagger$ 
Department of Physics and Winnipeg Institute of
Theoretical Physics,\\ 
University of Winnipeg\\
Winnipeg, Manitoba, Canada R3B 2E9\\
{[e-mail: g.kunstatter@uwinnipeg.ca]}\\[5pt]
}
{\sl
${}^\sharp$ 
School of Mathematical Sciences,
University of Nottingham\\
Nottingham NG7 2RD, United Kingdom\\
{[e-mail: jorma.louko@nottingham.ac.uk]}\\ [5pt]
}
{\sl
${}^\dagger$ 
Department of Physics and Astronomy, University of Manitoba\\
Winnipeg, Manitoba, Canada R3T 2N2\\
{[e-mail: j.ziprick-ra@uwinnipeg.ca]}\\[10pt]
}
\vspace{2ex}

{\bf Abstract}
\end{center}
We present a polymer quantization of the $-\lambda/r^2$ potential on
the positive real line and
compute numerically the bound state eigenenergies in terms of the
dimensionless coupling constant~$\lambda$. The singularity at the
origin is handled in two ways: first, by regularizing the potential
and adopting either symmetric or antisymmetric boundary conditions;
second, by keeping the potential unregularized but allowing the
singularity to be balanced by an antisymmetric boundary condition.
The results are compared to the semiclassical
limit of the polymer theory and to the conventional Schr\"odinger
quantization on $L_2(\mathbb{R}_+)$. The various quantization schemes
are in
excellent agreement for the highly excited states but differ for the
low-lying states, and the polymer spectrum is bounded below even when
the Schr\"odinger spectrum is not. We find as expected that for the
antisymmetric boundary condition the regularization of the potential
is redundant: the polymer quantum theory is well defined even with the
unregularized potential and the regularization of the potential does
not significantly affect the spectrum.
\vfill \hfill 
Revised December 2008

\vspace*{1ex}

\noindent
Published in
Phys.\ Rev.\ A {\bf 79}, 032104 (2009)

\end{titlepage}

\section{Introduction}

One of the most important outcomes expected of a successful theory of
quantum gravity is a clear and unambiguous solution to the problems
associated with the curvature singularities that are predicted by
classical general relativity. This expectation is natural since
quantum mechanics is known to cure classical singularities in other
contexts, such as the hydrogen atom.

In recent years there has been much work suggesting that Loop Quantum
Gravity (LQG) \cite{Thiemann07} may indeed resolve gravitational
singularities at least in the case of symmetry-reduced models, such
as spatially homogeneous \cite{cosm sing res} and inhomogeneous
\cite{MartinBenito:2008ej} cosmologies and spherically symmetric black
holes \cite{husain06,Ashtekar:2005qt,bh-interior}.  
Given the simplifications that
these models entail, it is pertinent to ask which features of the LQG
quantization scheme are crucial to the observed singularity
resolution.

There are two distinct, but related, features of the LQG quantization
program that appear to play a role in achieving singularity
resolution. The first is the fundamental discreteness that underlies
LQG due to its focus on holonomies of connections and associated
graphs embedded in a spatial manifold~\cite{Thiemann}. An analogous
approach in a purely quantum mechanical context is so-called polymer
quantization~\cite{AFW,halvorson}, in which the Hamiltonian dynamics
occurs on a discrete spatial lattice and the basic observables are the
operators associated with location on the lattice and translation
between lattice points. Polymer quantization provides a quantization
scheme that is mathematically and physically distinct from
Schr\"odinger quantization.

The second apparently key ingredient in the LQG singularity resolution
mechanism is the regularization of the singular terms in the
Hamiltonian using a trick first introduced in this context by
Thiemann~\cite{Thiemann}. The regularization is achieved by first
writing a classical inverse triad as the (singular) Poisson bracket of
classical phase space functions whose quantum counterparts are known,
and then defining the inverse triad operator as the commutator of
these quantum counterparts.  When applied to simple models this
procedure gives rise to quantum operators with bounded spectra. The
singularity is therefore kinematically ``removed'' from the spectra of
relevant physical operators, such as the inverse scale factor.

One question that arises concerns the role or perhaps the necessity of
the Thiemann trick in singularity resolution in LQG. Recall that in
the case of the hydrogen atom the singularity resolution is achieved
by defining self-adjoint operators in a Hilbert space. This requires a
careful choice of boundary conditions on the wave function
\cite{fewster-hydrogen} but does not require modification of the
singular $1/r$ potential. An example more relevant to quantum gravity
is the reduced Schr\"odinger quantization of the ``throat dynamics''
of the Schwarzschild interior, which on imposition of suitable boundary
conditions produced a discrete, 
bounded-from-below spectrum for the black hole
mass~\cite{louko96}.

Polymer quantization of the hydrogen atom was recently examined
in~\cite{HLW}, retaining only the s-wave sector and regularizing the
$1/r$ potential in a way that lets $r$ take values on the whole real
axis.  The choice of symmetric versus
antisymmetric boundary conditions at the singularity 
was found to have a signficant
effect on the ground state even after the
singularity itself had been regularized. 
In particular, in the limit of
small lattice separation the ground state eigenenergy showed evidence
of convergence towards the ground state energy of the
conventionally-quantized Schr\"odinger theory only for the 
antisymmetric boundary condition. 

In the present paper, we perform a similar polymer quantization of the
more singular $1/r^2$ potential.  When the potential is regularized,
we shall find that the choice of the boundary condition again has a
significant effect on the lowest-lying eigenenergies.  However, we
shall also find that the polymer theory with the antisymmetric
boundary condition is well defined even without regularizing the
potential, and with this boundary condition the regularized and
unregularized potentials yield closely similar spectra.  The boundary
condition at the singularity is hence not only a central piece of
input in polymer quantization, but it can even provide, along with the
modification to the kinetic term, the pivotal singularity avoidance
mechanism. While this is expected from general arguments that we will
make explicit later on, it is interesting and reassuring to see the
mechanism work in the special case of the $1/r^2$ potential, whose
degree of divergence is just at the threshold where a conventional
Schr\"odinger quantization will necessarily result into a spectrum
that is unbounded below. The polymer treatment of this system is thus
turning a Hamiltonian that is unbounded below into one that has a
well-defined ground state.

The $1/r^2$ potential is interesting in its own right: it has a
classical scale invariance that is broken by the quantum theory. In
addition, this potential appears frequently in black hole physics, for
example in the near horizon and near singularity behaviour of the
quasinormal mode potential~\cite{motl-quasi1,gk-quasi1}, in the near
horizon behaviour of scalar field propagation \cite{camblong-bh} and
in the Hamiltonian constraint in Painlev\'e-Gullstrand
coordinates~\cite{GKS}. It may therefore conceivably be of direct
relevance to quantum gravity.  There is a substantial literature on
Schr\"odinger quantization of this potential in $L_2(\mathbb{R}_+)$
(see for example
\cite{case,frank-potential,narnhofer,Gupta:1993id,camblong00,%
ordonez,inequivalence} and
the references in~\cite{fulop07}), but we are not aware of previous
work on polymer quantization of this potential.

Our paper is organized as follows. In Section \ref{sec:schrodinger
quantization} we review the Schr\"odinger quantization of the $1/r^2$
porential in $L_2(\mathbb{R}_+)$.  In Section \ref{sec:Polymer
Quantization} we formulate the polymer quantization of this system on
a lattice of fixed size and describe the numerical method. We also
include in this section a computation of the semi-classical polymer
spectrum from the Bohr-Sommerfeld quantization condition, with a fixed 
polymerization length scale.  
The numerical results are presented in Section~\ref{sec:results} 
and the conclusions are collected in Section~\ref{sec:conclusions}.

\section{Schr\"odinger quantization} 
\label{sec:schrodinger quantization}

We consider the classical Hamiltonian 
\be
H = p^2 - \frac{\lambda}{r^2} , 
\label{eq:H-classical}
\ee
where the phase space is $(r,p)$ with $r>0$ and
$\lambda\in\mathbb{R}$ is a constant.  We shall take $r$, $p$ and
$\lambda$ all dimensionless, and on quantization we set $\hbar=1$. 
If physical dimensions are restored, $r$
and $p$ will become expressed in terms of a single dimensionful scale
but $\lambda$ remains dimensionless.  That the coupling constant is
dimensionless is the speciality of a scale invariant potential.

Quantization of $H$ \nr{eq:H-classical} is of course subject to the
usual ambiguities. In particular, if one views $H$ as an effective
Hamiltonian that comes from a higher-dimensional configuration space
via symmetry reduction, with $r$ being a radial configuration
variable, the appropriate Hilbert space
may be 
$L_2(\mathbb{R}_+; m(r)dr)$, where $m$ is a positive-valued weight function. 
If for example 
$m(r) = r^a$, 
where $a\in\mathbb{R}$, 
then the ordering 
\be 
{\widehat H} = - \left(\frac{\partial^2}{\partial
r^2}+\frac{a}{r}\frac{\partial}{\partial r} \right) - \frac{\lambda}{r^2} 
\label{hamiltonian1}
\ee
makes the quantum Hamiltonian ${\widehat H}$ symmetric. 
If the wave function in $L_2(\mathbb{R}_+; m(r)dr)$ is denoted by~$\psi$, 
we may map $\psi$  to $\widetilde\psi \in L_2(\mathbb{R}_+; dr)$ by 
$\widetilde\psi(r) = r^{a/2} \psi(r)$, and ${\widehat H}$ is then mapped
in $L_2(\mathbb{R}_+; dr)$ to the Hamiltonian 
\be 
{\widehat {\widetilde H}} = - \frac{\partial^2}{\partial
r^2} - \frac{\widetilde\lambda}{r^2} , 
\ee 
where 
\be
\widetilde\lambda :=  
\lambda - \frac{a}{2}\left(\frac{a}{2}-1\right) . 
\ee
We shall consider any such mappings to have been done 
and take the quantum Hamiltonian to be simply 
\be 
{\widehat H} = - \frac{\partial^2}{\partial
r^2} - \frac{\lambda}{r^2}  , 
\label{quantum hamiltonian}
\ee 
acting in the Hilbert space $L_2(\mathbb{R}_+; dr)$.  

To guarantee
that the time evolution generated by ${\widehat H}$ 
\nr{quantum hamiltonian} is unitary, ${\widehat H}$
must be specified as a self-adjoint 
operator on $L_2(\mathbb{R}_+; dr)$~\cite{thirring-quantumbook}.  
A~comprehensive analysis of how to do this was given 
in \cite{narnhofer} 
(see also 
\cite{case,frank-potential,Gupta:1993id,camblong00,%
ordonez,inequivalence}). 
We shall review the results of \cite{narnhofer} in a way that
displays the spectrum explicitly 
for all the qualitatively different ranges of~$\lambda$. 

Before proceeding, we mention that several recent quantizations of the
$1/r^2$ potential first regularize the potential using various
renormalization techniques \cite{Gupta:1993id,camblong00,ordonez}. 
In particular, when the
spectrum of a self-adjoint extension is unbounded below, these
renormalization techniques need not lead to an equivalent quantum
theory~\cite{inequivalence}. We shall here discuss only the
self-adjoint extensions.

To begin, recall \cite{thirring-quantumbook} 
that the deficiency indices of ${\widehat H}$ are found by considering 
normalizable solutions to the eigenvalue equation 
${\widehat H}\psi = \pm i \psi$. An elementary analysis shows that 
${\widehat H}$ is essentially
self-adjoint for $\lambda \le - 3/4$, 
but for $\lambda > - 3/4$ 
a boundary condition at $r=0$ is 
needed to make ${\widehat H}$ self-adjoint. 
Physically, this boundary condition will ensure that no
probability is flowing in/out at $r=0$.

\subsection{$\lambda> 1/4$}

For $\lambda > 1/4$, we write 
$\lambda = 1/4 + \alpha^2$ with $\alpha>0$. 

For $E>0$, the 
linearly independent (non-normalizable) 
solutions to the eigenvalue equation 
\be
{\widehat H}\psi = E \psi
\label{eq:eigenvalue}
\ee 
are 
$\sqrt{r} \, J_{\pm i \alpha}(\sqrt{E} \, r)$.  
These oscillate infinitely many times as $r\to0$. To find the boundary
condition, we consider the linear combinations
\be 
\psi_E(r) := \sqrt{r} \left[ e^{i\beta}
E^{-i\alpha/2} J_{i \alpha}(\sqrt{E} \, r) + e^{-i\beta} E^{i\alpha/2}
J_{-i \alpha}(\sqrt{E} \, r) \right] , 
\label{eq:psi_E}
\ee 
where $\beta$ is a parameter that a priori could depend on~$E$. 
As $\psi_E$ is periodic in $\beta$ with period~$2\pi$, 
and as replacing 
$\beta$ by $\beta + \pi$ multiplies $\psi_E$ by~$-1$, 
we may understand $\beta$ periodic with period~$\pi$. 
For concreteness, we
could choose for example $\beta \in [0, \pi)$.

For the probability flux through $r=0$ to vanish, we need 
\be 
\overline{\psi_E} \, \partial_r \psi_{E'} - \overline{\psi_{E'}}
\, \partial_r \psi_{E} \to 0 \ \ \text{as $r\to0$} 
\ee 
for all $E$ and~$E'$, 
where the overline denotes the complex conjugate. 
Using the small argument asymptotic form (equation (9.1.7) in~\cite{AS}) 
\be 
J_{\nu}(z)\to\frac{(z/2)^\nu}{\Gamma(\nu+1)} \ \ \text{as $z\to0$},
\ee
this is seen to require $\sin(\beta-\beta')=0$, and hence
$\beta$ must be independent of~$E$. The choice of the constant $\beta$ hence
specifies the boundary condition at the origin.

To find the eigenvalues, we consider the normalizable solutions
to~\nr{eq:eigenvalue}. Such solutions exist only when $E<0$, and they
are $\sqrt{r} \, K_{i \alpha}(\sqrt{-E} \, r)$.  These solutions must
satisfy at $r\to0$ the same boundary condition as
$\psi_E$~\nr{eq:psi_E}. Using the small argument asymptotic form
(equations (9.6.2) and (9.6.7) in~\cite{AS}) \be
K_\nu(Z)=K_{-\nu}(z)\to \frac{\pi}{\sin(\nu\pi)}\left[
\frac{(z/2)^{-\nu}}{\Gamma(-\nu+1)}-\frac{(z/2)^\nu}{\Gamma(\nu+1)}\right]
\ \ \text{as $z\to0$}, \ee this shows that the eigenvalues are \be E_n
= E_0 \exp(- 2\pi n/\alpha) , \ \ n \in \mathbb{Z} ,
\label{tower}
\ee
where 
\be
E_0 = - \exp[(2\beta +\pi)/\alpha] . 
\label{E0}
\ee
This spectrum 
is an infinite tower, with $E_n \to 0_-$ as $n\to\infty$
and $E_n \to -\infty$ as $n\to-\infty$.  The spectrum is unbounded
from below.

We note that Schr\"odinger quantization of a regulated 
form of the potential 
yields a semi-infinite tower of states that is 
similar to \nr{tower} as $n\to\infty$ 
but has a ground state~\cite{camblong00}.  
The energy of the ground state goes to 
$-\infty$ when the regulator is removed.

\subsection{$\lambda = 1/4$}

For $\lambda = 1/4$, the 
solutions to the eigenvalue equation 
\nr{eq:eigenvalue}
for $E>0$ are 
$\sqrt{r} \, J_0(\sqrt{E} \, r)$ and 
$\sqrt{r} \, N_0(\sqrt{E} \, r)$. 
We consider the linear combinations 
\be 
\psi_E(r) := \sqrt{r} \left\{ (\cos\beta)
J_0(\sqrt{E} \, r) 
+ (\sin\beta)
\left[ 
\frac\pi2 N_0 (\sqrt{E} \, r) 
- \ln\left( \frac{\sqrt{E} \, e^\gamma}{2}\right)
J_0 (\sqrt{E} \, r) 
\right]
\right\} , 
\label{eq:psi_E-m}
\ee 
where $\gamma$ is Euler's constant and $\beta$ is again a parameter
that may be understood periodic with period $\pi$ and could a priori
depend on~$E$.  As above, we find that $\beta$ must be a constant
independent of $E$ and its value determines the boundary condition at
the origin.

Normalizable solutions to \nr{eq:eigenvalue} exist only for
$E<0$. They are $\sqrt{r} \, K_0(\sqrt{-E} \, r)$, and they must
satisfy the same boundary condition as $\psi_E$ \nr{eq:psi_E-m} at
$r\to0$. Using the small argument expansion of $K_0$~\cite{AS}, we
find that for $\beta=0$ there are no normalizable states, while for $0
< \beta < \pi$ there is exactly one normalizable state, with the
energy
\be
E_0 = - 4 \exp( -2\gamma + 2 \cot\beta) . 
\ee

\subsection{$-3/4 < \lambda < 1/4$} 

For $-3/4 < \lambda < 1/4$, we write 
$\lambda = 1/4 - \nu^2$ with $0<\nu<1$. 

The solutions to the eigenvalue equation 
\nr{eq:eigenvalue}
for $E>0$ are 
$\sqrt{r} \, J_{\pm \nu}(\sqrt{E} \, r)$. 
Considering the linear combinations 
\be 
\psi_E(r) := \sqrt{r} \left[ (\cos\beta)
E^{-\nu/2} J_{\nu}(\sqrt{E} \, r) + (\sin\beta) E^{\nu/2}
J_{-\nu}(\sqrt{E} \, r) \right] , 
\label{eq:psi_E-nu}
\ee 
we find as above that $\beta$ is a constant, understood periodic with
period~$\pi$, and its value specifies the boundary condition at the
origin.

Normalizable solutions to \nr{eq:eigenvalue} exist only for 
$E<0$. They are 
$\sqrt{r} \, K_{\nu}(\sqrt{-E} \, r)$ 
and must satisfy the same boundary condition as $\psi_E$ \nr{eq:psi_E-nu}
at $r\to0$. 
Using the small argument asymptotic form of $K_{\nu}$~\cite{AS}, we
find that there are no normalizable states for $0\le \beta \le \pi/2$,
while for $\pi/2 < \beta < \pi$ there is exactly one normalizable
state, with the energy
\be
E_0 = - {(- \cot\beta)}^{1/\nu} . 
\ee

We note that in the special case of a free particle, $\lambda=0$, 
the Bessel functions reduce to 
trigonometric and exponential functions.

\subsection{$\lambda \le -3/4$} 

For $\lambda \le -3/4$, we write $\lambda = 1/4 - \nu^2$ with $\nu \ge
1$. ${\widehat H}$~is now essentially self-adjoint. Any prospective
normalizable solution to \nr{eq:eigenvalue} would need to have $E<0$
and take the form $\sqrt{r} \, K_{\nu}(\sqrt{-E} \, r)$, but since now
$\nu\ge1$, these solutions are not normalizable and hence do not
exist.

\section{Polymer quantization}
\label{sec:Polymer Quantization}

In this section we develop the polymer quantization of the $1/r^2$
potential We proceed as in~\cite{HLW}, briefly reiterating the main
steps for completeness.

It is necessary to extend the $r$ coordinate to negative values with
the replacement $r \rightarrow x \in \mathbb{R}$ in order to use
central finite difference schemes at the origin.  This will allow us
to introduce at the origin both a symmetric boundary condition (with
the regulated potential developed in subsection~\ref{subsec:fqpt}) and
an antisymmetric boundary condition.

The polymer Hilbert space on the full real line is spanned by the
basis states
\begin{equation}
\psi_{x_0}(x) = \left\{
	\begin{array}{ll}
	1, & x=x_0\\
	0, & x\neq x_0
	\end{array}
\right.
\end{equation}
with the inner product
\begin{equation}
(\psi_x , \psi_{x^\prime}) = \delta_{x,x^\prime},
\label{eq:bohr-ip}
\end{equation}
where the object on the right hand side is the Kronecker delta.
The position operator acts by multiplication as
\begin{equation}
(\hat{x} \psi)(x) = x \psi(x).
\label{xact}
\end{equation}
Defining a momentum operator takes more
care. Consider the translation operator~$\widehat{U}$, which acts
as
\begin{equation}
(\widehat{U}_\mu \psi)(x) = \psi(x+\mu). 
\label{Uact}
\end{equation}
In ordinary Schr\"odinger quantization we would have
$\widehat{U}_\mu=e^{i\mu \hat{p}}$.  Following~\cite{AFW}, we hence
define the momentum operator and its square as
\begin{subequations} 
\begin{eqnarray}
\hat{p} &=& \frac{1}{2i\mmuu}(\widehat{U}_\mmuu - \widehat{U}_\mmuu^\dagger),\\
\widehat{p^2} &=& \frac{1}{\mu^2}(2 - \widehat{U}_\mu - \widehat{U}_\mu^\dagger),
\end{eqnarray}
\end{subequations}
where $\mmuu := \mu/2$. 
We may thus write the polymer Hamiltonian as
\begin{equation}
\widehat{H}_{\mathrm{pol}} = 
\widehat{T}_{\mathrm{pol}} 
+ 
\widehat{V}_{\mathrm{pol}} \ , 
\label{Hpol}
\end{equation}
where 
\begin{subequations}
\begin{eqnarray}
\widehat{T}_{\mathrm{pol}} 
&=& 
\frac{1}{\mu^2}
(2 - \widehat{U}_\mu - \widehat{U}_\mu^\dagger), 
\label{Tpol}
\\
\widehat{V}_{\mathrm{pol}} 
&=&
- \frac{\lambda}{{\hat{x}^2}}.
\label{Vpol}
\end{eqnarray}
\end{subequations}

Considering the action of $\hat{x}$ and $\widehat{U}_\mu$, we see that
the dynamics generated by (\ref{Hpol}) separates the polymer Hilbert
space into an infinite number of superselection sectors, each having
support on a regular $\mu$-spaced lattice \{$\Delta + n\mu$ $|$ $n \in
\mathbb{Z}$\}. The choice of \{$\Delta$ $|$ $0 \leq \Delta < \mu$\}
picks the sector. Since we wish to study singularity resolution, we
concentrate on the $\Delta=0$ sector, which we expect the singularity
of the potential to affect most. We shall discuss the regularization
of $\widehat{V}_{\mathrm{pol}}$ at  
this singularity in subsection~\ref{subsec:fqpt}.

\subsection{Semiclassical polymer theory} 

Before analyzing the full polymer quantum theory, we examine the
semiclassical polymer spectrum using the Bohr-Sommerfeld
quantization condition.

Following 
\cite{cosm sing res,Ashtekar:2005qt,bh-interior}, 
we take the classical limit
of the polymer Hamiltonian (\ref{Hpol}) by 
keeping the polymerization scale $\mu$ fixed and 
making the replacement 
$\widehat{U}_\mu \to e^{i\mu p}$, 
where $p$ is the classical momentum. Note that this is 
different from the continuum limit in which $\mu$ goes to
zero and the quantum theory is expected to be equivalent to
Schr\"odinger quantization \cite{corichi07}.

We assume $\lambda>0$. 
It follows, as will be verified below, 
that the classical polymer 
orbits never reach the origin, 
and we may hence assume the configuration variable $x$ to be 
positive and revert to the symbol~$r$. 
The classical polymer Hamiltonian takes thus the form 
\be
H_{\mathrm{pol}} = \frac{\sin^2(\mmuu p)}{\mmuu^2} - \frac{\lambda}{r^2} .
\label{eq:class-pol-ham}
\ee 
Note that $H_{\mathrm{pol}}$ reduces to the classical 
non-polymerized Hamiltonian \nr{eq:H-classical} 
in the limit $\mmuu\to0$. 

A first observation is that the 
kinetic term in 
$H_{\mathrm{pol}}$ 
is non-negative and bounded above by~$1/{\mmuu^2}$. 
Denoting the time-independent value of 
$H_{\mathrm{pol}}$ on a classical solution by~$E$, 
it follows that $E$ is bounded above by 
\be
E < E_{\mathrm{max}} := \frac{1}{\mmuu^2} , 
\ee
and on a given classical solution $r$ 
is bounded below by $r\ge r_-$, where 
\be
r_- := \left(\frac{\lambda \mmuu^2}{1-\mmuu^2 E }\right)^{1/2} . 
\ee

An elementary analysis shows every classical 
solution has a bounce at $r = r_-$. 
For $E\ge0$ this is the only turning point, and the solution 
is a scattering solution, 
with $r\to\infty$ as $t\to\pm\infty$. 
For $E<0$ there is a second turning point at $r=r_+ > r_-$, where 
\be
r_+ := \left(\frac{\lambda}{-E}\right)^{1/2} , 
\ee
and the solution is a bound solution, 
with $r$ oscillating periodically between 
$r_+$ and~$r_-$. 
Note that $r_+$ is independent of~$\mmuu$, 
and the outer turning point in fact coincides 
with the turning point of the non-polymerized classical theory. 

The classical polymer solutions are thus qualitatively 
similar to the classical non-polymerized 
solutions at large~$r$, both for $E\ge0$ and for $E<0$. 
What is different is that the polymer energy is bounded from above, 
and more importantly that the polymer solutions bounce at $r=r_-$. 
In this sense the classical polymer theory has resolved the 
singularity at $r=0$. 
The resolution depends on the polymerization scale: 
for fixed~$E$, $r_- = \mmuu \sqrt{\lambda}
\bigl[ 1 + O(\mmuu^2) \bigr] 
\to 0$ as $\mmuu\to0$, and for fixed~$\mmuu$, 
$r_- 
\to \mmuu \sqrt{\lambda}$ as $E\to0$. 


As the $E<0$ solutions are periodic, we can use 
the Bohr-Sommerfeld quantization condition 
to estimate the semiclassical quantum spectrum. 
A~subtlety here is that semiclassical estimates already in 
ordinary Schr\"odinger quantization with a $1/r^2$ term involve a
shift in the coefficient of this term~\cite{AH}. Anticipating a
similar shift here, we look at the Bohr-Sommerfeld quantization
condition with $\lambda$ replaced by~$\leff$, and we will then
determine $\leff$ by comparison with the Schr\"odinger quantization.

For a classical solution with given~$E$, formula 
\nr{eq:class-pol-ham} implies (with $\lambda$ replaced by~$\leff$) 
\be
r = \frac{\mmuu \sqrt\leff}{\sqrt{ \sin^2(\mmuu p) - \mmuu^2 E}}.
\ee
Taking $E<0$, 
the phase space integral $J(E) := \oint r\,dp$ 
over a full cycle is hence 
\begin{align}
J(E) &= 
\oint r\,dp
\nonumber
\\
&= 
2 \sqrt\leff \int_0^{\pi/(2\mmuu)}
\frac{\mmuu \, dp}{\sqrt{ \sin^2(\mmuu p) - \mmuu^2 E}}
\qquad (p = y/\mmuu) 
\nonumber
\\
&= 
2 \sqrt\leff \int_0^{\pi/2}
\frac{dy}{\sqrt{ \sin^2(y) - \mmuu^2 E}} . 
\nonumber
\\
&= 
\frac{2 \sqrt\leff}{\sqrt{1 - \mmuu^2 E}}
\int_0^{\pi/2}
\frac{dy}{\sqrt{1 - {(1 - \mmuu^2 E)}^{-1} \cos^2(y)}} 
\nonumber
\\
&= 
\frac{2 \sqrt\leff}{\sqrt{1 - \mmuu^2 E}}
\, K \left( {(1 - \mmuu^2 E)}^{-1/2} \right) , 
\label{eq:J-finalform}
\end{align} 
where $K$ is the complete elliptic integral 
of the first kind~\cite{gradstein}. 
In the limit $\mmuu^2 E \to 0$, the expansion 
(8.113.3) in \cite{gradstein} yields  
\begin{align}
J(E) &= 
2 \sqrt\leff 
\left[ 
\ln\left(
\frac{4}{\mmuu\sqrt{-E}}
\right)
+ \mathbb{O}\left(\mmuu\sqrt{-E}\right)
\right] . 
\label{eq:J-asymptotic}
\end{align} 
The Bohr-Sommerfeld quantization condition now 
states that the eigenenergies of the highly 
excited states are given asymptotically by $J(E) = 2\pi n$, 
where $n \gg 1$ is an integer. 
By~\nr{eq:J-asymptotic}, this gives the asymptotic eigenenergies 
\be
E_n = - \frac{16}{\mmuu^2} \exp \bigl (- 2\pi n/\sqrt\leff  \bigr) , 
\ \ n\to\infty . 
\label{eq:pol-WKB-evs}
\ee

The Bohr-Sommerfeld estimate \nr{eq:pol-WKB-evs} agrees with 
the spectrum \nr{tower} obtained from conventional 
Schr\"odinger quantization 
for $\lambda > 1/4$, 
provided $\leff = \lambda - \frac14$ 
and we choose in \nr{tower} the self-adjoint extension for which 
\be
\beta = - \frac{\pi}{2} + \alpha \ln(4/\mmuu)  
\ \ \ 
(\text{mod $\pi$}) .  
\label{eq:BS-beta}
\ee The shift $\leff = \lambda - \frac14$ is exactly that which arises
in ordinary Schr\"odinger quantization of potentials that include a
$1/r^2$ term: the reason there is the matching of the small $r$
behaviour of the exact eigenstates to the WKB approximation.  For a
lucid analysis of this phenomenon in the quasinormal mode context, see
the discussion between equations (23) and (28) in~\cite{AH}.  Note,
however, that in our system the Bohr-Sommerfeld condition cannot be
applied directly to the unpolymerized theory, since $J(E)$
\nr{eq:J-finalform} diverges as $\mmuu\to0$.

\subsection{Full quantum polymer theory}
\label{subsec:fqpt}

We now return to the full polymer quantum theory, 
with the Hamiltonian \nr{Hpol} and $\lambda \in\BbbR$. 

We write the basis states in Dirac notation as $\left| m\mu
\right\rangle$, where $m\in\BbbZ$. Writing a state 
in this basis as 
$\psi = \sum_m c_m
\left|m\mu\right\rangle$, it follows from 
\nr{eq:bohr-ip} 
that the inner product reads 
$\left( \psi^{(1)}, \psi^{(2)} \right) = 
\sum_m \overline{{c_m}^{(1)}} \, c_m^{(2)}$. 
The Hilbert space is thus~$L_2(\BbbZ)$. 
It will be useful to decompose this Hilbert space as the direct sum 
$L_2(\BbbZ) = L_2^s(\BbbZ) \oplus L_2^a(\BbbZ)$, where the states in
the symmetric sector $L_2^s(\BbbZ)$
satisfy $c_m = c_{-m}$ and the states in the antisymmetric sector 
$L_2^a(\BbbZ)$
satisfy $c_m = - c_{-m}$. 

The action of $\widehat{T}_{\mathrm{pol}}$ \nr{Tpol} reads 
\begin{equation}
\widehat{T}_{\mathrm{pol}} 
\left( 
\sum_m c_m \left|m\mu\right\rangle 
\right) 
= \frac{1}{\mu^2}
\sum_m 
\left( 
2 c_m - c_{m+1} - c_{m-1} 
\right) 
\left|m\mu\right\rangle . 
\label{eq:T-action}
\end{equation}
$\widehat{T}_{\mathrm{pol}}$ 
is clearly a bounded operator on $L_2(\BbbZ)$. 
$\widehat{T}_{\mathrm{pol}}$ is manifestly symmetric, and 
an explicit solution of the eigenvalue equation 
$\widehat{T}_{\mathrm{pol}} \psi = E \psi$, given in equation 
\nr{recsol} below,  
shows that there are no normalizable solutions with $E=\pm i$. 
$\widehat{T}_{\mathrm{pol}}$ is hence essentially 
self-adjoint (\cite{reed-simonII}, Theorem X.2). 
It is also positive, since 
$\left( \psi, \widehat{T}_{\mathrm{pol}}\psi \right) > 0$ 
for any $\psi \ne 0$ by the Cauchy-Schwarz inequality. 

The action of $\widehat{V}_{\mathrm{pol}}$ \nr{Vpol} reads
\begin{equation}
\widehat{V}_{\mathrm{pol}} 
\left( 
\sum_m c_m \left|m\mu\right\rangle 
\right) 
= -\frac{\lambda}{\mu^2}
\sum_m 
f_m^{\mathrm{pol}} \, c_m
\left|m\mu\right\rangle , 
\label{eq:V-action}
\end{equation}
where 
\begin{equation}
f_m^{\mathrm{pol}} := \frac{1}{m^2}. 
\label{eq:fpol-def}
\end{equation}
As \nr{eq:V-action} is 
ill-defined on any 
state for which $c_m\ne0$, 
$\widehat{V}_{\mathrm{pol}}$ 
is not a densely-defined operator on $L_2(\BbbZ)$. 
We consider two ways to handle this singularity. 

The first way is to regulate $\widehat{V}_{\mathrm{pol}}$ explicitly. 
Recall that for $x \in \BbbR\setminus \{0\}$ 
we can write  
\begin{equation}
\frac{\sgn(x)}{\sqrt{|x|}} = 2 \frac{d(\sqrt{|x|})}{dx} , 
\label{divsub}
\end{equation}
and on our lattice this can be implemented as 
the finite difference expression 
\begin{equation}
\frac{\sgn(x)}{\sqrt{|x|}} \rightarrow \frac{1}{\mu}
\left(\sqrt{|x_{m+1}|} - \sqrt{|x_{m-1}|} \right) + O(\mu^2).
\label{eq:reg-invsqrt}
\end{equation}
We hence define the regulated polymer version of 
$\sgn(x)/\sqrt{|x|}$ by dropping the 
$O(\mu^2)$ term in~\nr{eq:reg-invsqrt}, and we define the 
regulated polymer potential 
$\widehat{V}_{\mathrm{pol}}^{\mathrm{reg}}$ 
by raising this to the fourth power, 
\begin{equation}
\frac{\lambda}{{(x_m)}^2} \rightarrow 
\frac{\lambda}{\mu^4}
\left(\sqrt{|x_{m+1}|} - \sqrt{|x_{m-1}|} \right)^4 , 
\end{equation}
or 
\begin{equation}
\widehat{V}_{\mathrm{pol}}^{\mathrm{reg}}
\left( 
\sum_m c_m \left|m\mu\right\rangle 
\right) 
= - \frac{\lambda}{\mu^2} 
\sum_m f_m^{\mathrm{reg}} \, c_m 
\left|m\mu\right\rangle , 
\label{eq:Vreg-action}
\end{equation}
where 
\begin{equation}
f_m^{\mathrm{reg}} := 
\left(\sqrt{|m+1|} - \sqrt{|m-1|} \right)^4.
\label{eq:freg-def}
\end{equation}
$\widehat{V}_{\mathrm{pol}}^{\mathrm{reg}}$ is clearly 
a bounded essentially self-adjoint operator on $L_2(\BbbZ)$, 
and its operator norm is 
$4|\lambda|/(\mu^2)$. 

The regulated polymer Hamiltonian can now be defined by 
\begin{equation}
\widehat{H}_{\mathrm{pol}}^{\mathrm{reg}}
= 
\widehat{T}_{\mathrm{pol}}
+ 
\widehat{V}_{\mathrm{pol}}^{\mathrm{reg}} . 
\end{equation}
It follows by the 
Kato-Rellich theorem 
(\cite{reed-simonII}, Theorem X.12) 
that $\widehat{H}_{\mathrm{pol}}^{\mathrm{reg}}$ 
is essentially self-adjoint on $L_2(\BbbZ)$ 
and bounded below by $-4|\lambda|/(\mu^2)$. 
Further, both $\widehat{T}_{\mathrm{pol}}$ 
and 
$\widehat{V}_{\mathrm{pol}}^{\mathrm{reg}}$ leave
$L_2^s(\BbbZ)$ and
$L_2^a(\BbbZ)$ invariant, and their
boundedness and self-adjointness properties 
mentioned above hold also for their 
restrictions to $L_2^s(\BbbZ)$ and
$L_2^a(\BbbZ)$. It follows that 
$\widehat{H}_{\mathrm{pol}}^{\mathrm{reg}}$ 
restricts to both 
$L_2^s(\BbbZ)$ and
$L_2^a(\BbbZ)$ 
as a self-adjoint 
operator bounded below by $-4|\lambda|/(\mu^2)$. 
We denote both of these restrictions by 
$\widehat{H}_{\mathrm{pol}}^{\mathrm{reg}}$, 
leaving the domain to be understood from
the context. 

The second way to handle the singularity of 
$\widehat{V}_{\mathrm{pol}}$ \nr{eq:V-action}
is to restrict at the outset to 
the antisymmetric subspace $L_2^a(\BbbZ)$, where 
$\widehat{V}_{\mathrm{pol}}$ is 
essentially self-adjoint and its 
operator norm is $|\lambda|/(\mu^2)$. 
It follows as above that the unregulated polymer Hamiltonian 
\begin{equation}
\widehat{H}_{\mathrm{pol}}
= 
\widehat{T}_{\mathrm{pol}}
+ 
\widehat{V}_{\mathrm{pol}}
\end{equation}
on $L_2^a(\BbbZ)$
is essentially self-adjoint and bounded below 
by $-|\lambda|/(\mu^2)$. 

Two comments are in order. First, 
$\widehat{H}_{\mathrm{reg}}$ can be written in terms of operators as 
\begin{equation}
\widehat{H}_{\mathrm{reg}} 
= \frac{1}{\mu^2}(2 - \widehat{U}_\mu - \widehat{U}_\mu^\dagger) 
- \frac{\lambda}{\mu^4}\left( \widehat{U}_\mu \sqrt{|\hat{x}|} 
\widehat{U}_\mu^\dagger - \widehat{U}_\mu^\dagger \sqrt{|\hat{x}|} 
\widehat{U}_\mu \right)^4.
\label{Hreg}
\end{equation}
The potential in (\ref{Hreg}) can hence 
be viewed as arising by the substitution 
\begin{equation}
\frac{\sgn(x)}{\sqrt{|x|}} 
\rightarrow \frac{2}{i\mu}\widehat{U}_\mu^\dagger 
\left\{ \sqrt{|x|},\widehat{U}_\mu \right\} , 
\end{equation}
in place of~(\ref{divsub}). 
This method is similar to Thiemann's
regularization of inverse triad operators 
in loop quantum gravity~\cite{Thiemann}.

\begin{figure}
\begin{center}
\includegraphics[width=0.75\linewidth]{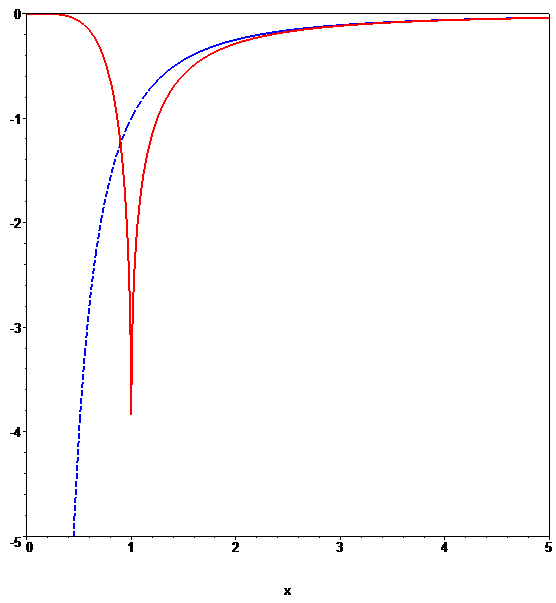}
\caption{The solid (red) line is the 
regulated $-1/x^2$ potential 
($\lambda=1$, $\mu=1$). 
The dashed (blue) line is the 
unregulated potential.}
\label{x2plot}
\end{center}
\end{figure}

Second, the regulated potential vanishes at the origin but 
is greater in absolute value than the unregulated potential for
$|m|\geq1$. However, the difference is significant only for the 
lowest few~$|m|$, and the two potentials quickly 
converge as $|m|\to\infty$, 
as shown in Figure~\ref{x2plot}. 
The regulated and unregulated potentials hence differ 
significantly only near the singularity.

\subsection{Eigenstates and the numerical method}
\label{sec:num-method}

We are now ready to look for the eigenstates 
of the Hamiltonian. 
Writing the eigenstate as 
$\psi = \sum_m c_m \left|m\mu\right\rangle$ 
and denoting the eigenvalue by~$E$, 
the regulated eigenvalue equation 
$\widehat{H}_{\mathrm{pol}}^{\mathrm{reg}} \psi = E \psi$ 
and the unregulated eigenvalue equation 
$\widehat{H}_{\mathrm{pol}} \psi = E \psi$ 
both give a recursion relation that takes the form 
\begin{equation}
c_m\left( 2 - \mu^2 E -\lambda f_m \right) = c_{m+1} + c_{m-1} , 
\label{recursion}
\end{equation}
where
$f_m = f_m^{\mathrm{reg}}$ 
\nr{eq:freg-def}
for the regulated potential and 
$f_m = f_m^{\mathrm{pol}}$ 
\nr{eq:fpol-def} for the unregulated potential. 
Note that the polymerization scale 
$\mu$ enters this recursion relation 
only in the
combination~$\mu^2 E$, whether or not the potential is regulated. 
This is a direct consequence of the scale
invariance of the potential. 

From now on, we take $\lambda>0$ and $E<0$. 

We use the ``shooting method'' that was applied in \cite{HLW} to the
polymerized $1/|x|$ potential. 
For large~$|m|$, (\ref{recursion}) is
approximated by
\begin{equation}
c_m\left( 2 - \mu^2 E \right) = c_{m+1} + c_{m-1}. 
\label{recapprox}
\end{equation}
The linearly independent solutions to \nr{recapprox}
are
\begin{equation}
c_m = \left[1 - \frac{\mu^2 E}{2} 
+ \sqrt{\left(1 - \frac{\mu^2 E}{2}\right)^2-1}\right]^{\pm m}. 
\label{recsol}
\end{equation}
The upper (respectively lower) sign
gives coefficients that increase (decrease)
exponentially as $m \rightarrow \infty$. We can therefore use
\nr{recsol} with the lower sign to set the initial conditions at large
positive $m$ \cite{elaydi}. 

To set up the shooting problem, we choose a value for $\mu^2 E$ and
begin with some $m_0 \gg \sqrt{\frac{\lambda}{\mu^2|E|}}$ to find
$c_{m_0}$ and $c_{m_0 - 1}$ using the approximation~(\ref{recsol}). We
then iterate downwards with~(\ref{recursion}). In the antisymmetric
sector, we stop the iteration at $c_0$ and shoot for values of $\mu^2
E$ for which $c_0=0$. This shooting problem is well defined both for
the unregulated potential \nr{eq:V-action} and for the regulated
potential~\nr{eq:Vreg-action}, since the computation of $c_0$ via
(\ref{recursion}) does not require evaluation of $f_m$ at $m=0$.  In
the symmetric sector, we stop at the iteration at $c_{-1}$ and shoot
for values of $\mu^2 E$ for which $c_{-1}=c_1$.  As the computation of
$c_{-1}$ requires evaluation of $f_m$ at $m=0$, the symmetric sector
is well defined only for the regulated potential.

\section{Results}
\label{sec:results}

We shall now compare the spectra of full polymer quantization,
Bohr-Sommerfeld polymer quantization and ordinary Schr\"{o}dinger
quantization. We are particularly interested in the sensitivity of
the results to the choice of the symmetric versus the antisymmetric
sector.

First of all, we find that when the potential is regulated, the choice
of the symmetric versus antisymmetric boundary condition in the full
polymer quantum theory has no significant qualitative effect for
sufficiently large~$\lambda$, the only difference being slightly lower
eigenvalues for the symmetric boundary condition.  The lowest five
eigenvalues in the two sectors are shown in Table \ref{evorder} for
$\lambda=2$.  This is in a sharp contrast with what was found in
\cite{HLW} for the $1/r$ potential, where the symmetric sector
contained a low-lying eigenvalue that appeared to tend to $-\infty$ as
the polymerization scale was decreased.

\begin{table}[htb]
\begin{center}
\begin{tabular}{|c|c|c|}
\hline
 & antisymmetric & symmetric\\
\hline
$E_0$&-6.14&-6.37\\
$E_1$&$-2.35\cdot10^{-2}$&$-2.43\cdot10^{-2}$\\
$E_2$&$-2.03\cdot10^{-4}$&$-2.10\cdot10^{-4}$\\
$E_3$&$-1.76\cdot10^{-6}$&$-1.82\cdot10^{-6}$\\
$E_4$&$-1.52\cdot10^{-8}$&$-1.57\cdot10^{-8}$\\
\hline
\end{tabular}
\end{center}
\caption{The lowest five eigenvalues of the regulated potential with 
antisymmetric and symmetric boundary 
conditions ($\lambda=2$, $\mu=1$).}
\label{evorder}
\end{table}

\begin{figure}
\begin{center}
\includegraphics[width=1.0\linewidth]{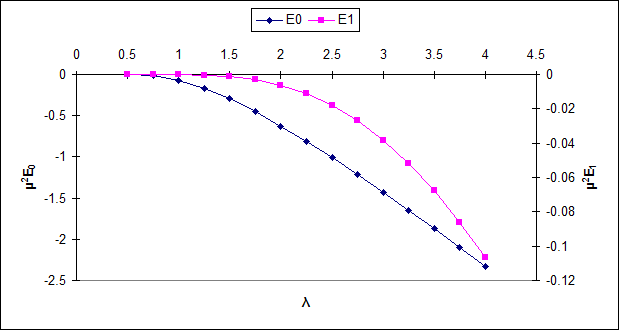}
\caption{The lowest two energy levels as a function of 
$\lambda$ for the unregulated potential with 
antisymmetric boundary conditions.}
\label{x2urEvsL}
\end{center}
\end{figure}

\begin{figure}
\begin{center}
\includegraphics[width=1.0\linewidth]{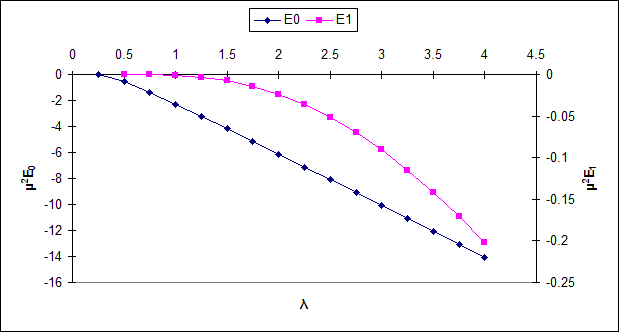}
\caption{The lowest two energy levels as a 
function of $\lambda$ for the regulated 
potential with antisymmetric boundary conditions.}
\label{x2rEvsL}
\end{center}
\end{figure}

Another key feature is that for sufficiently large $\lambda$ there is
indeed a negative energy ground state. For $3\le\lambda\le4$, the
plots of the lowest eigenvalues as a function of $\lambda$ in Figures
\ref{x2urEvsL} and \ref{x2rEvsL} show that the analytic lower bound
obtained in subsection \ref{subsec:fqpt} is accurate within a factor
of $1.2$ for the regulated potential in the antisymmetric sector and
within a factor of two for the unregulated potential.

Figures \ref{x2urEvsL} and \ref{x2rEvsL} also 
indicate that the lowest eigenvalues
converge towards zero as $\lambda$ decreases, for both the
unregulated and regulated potentials, with the unregulated eigenvalues
reaching zero slightly before the regulated. Near $E_n=0$ the
relationship is quadratic in $\lambda$ while the plots straighten out
to a linear relationship for larger~$\lambda$. 

The numerics become slow as the energies are close to zero.  We were
unable to investigate systematically whether bound states exist for
$\lambda\leq1/4$, and in particular to make a comparison with the
single bound state that occurs in Schr\"odinger quantization with
certain self-adjoint extensions.  For $\lambda$ slightly below~$1/4$,
we do find one bound state, but we do not know whether the absence of
further bound states is a genuine property of the system of an
artefact of insufficient computational power. This would be worthy of
further investigation. The eigenvalues show a similar dependence on
$\lambda$ for both regulated and unregulated potentials, with the
energies for the regulated potential being lower than those for the
unregulated version as one would expect from comparing the potentials
as in Figure~\ref{x2plot}.

\begin{figure}
\begin{center}
\includegraphics[width=0.75\linewidth]{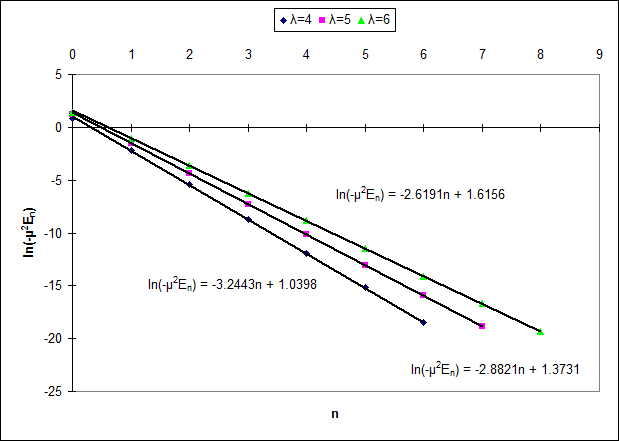}
\caption{$\ln(E_n)$ vs.\ $n$ for the unregulated 
potential with antisymmetric boundary conditions 
with linear fits ($R^2=1$).}
\label{x2urEvsn}
\end{center}
\end{figure}

\begin{figure}
\begin{center}
\includegraphics[width=0.75\linewidth]{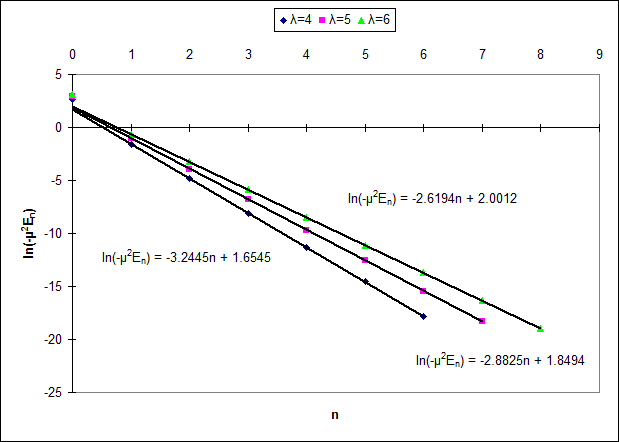}
\caption{$\ln(E_n)$ vs.\ $n$ for the regulated 
potential with antisymmetric boundary 
conditions with linear fits ($R^2=1$).}
\label{x2rEvsn}
\end{center}
\end{figure}

\begin{figure}
\begin{center}
\includegraphics[width=1.0\linewidth]{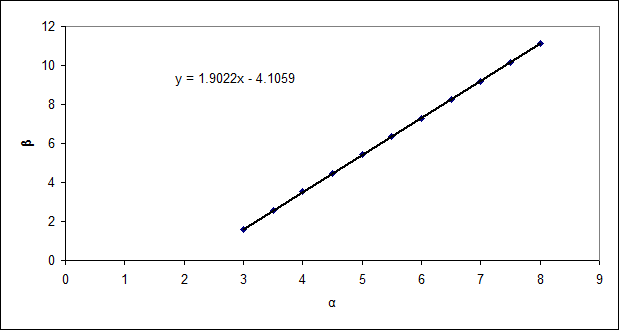}
\caption{$\beta$ vs.\ $\alpha$ with $\mu=1$ 
for the unregulated potential with antisymmetric 
boundary conditions with a linear fit ($R^2=1$).}
\label{x2urBvsA}
\end{center}
\end{figure}

\begin{figure}
\begin{center}
\includegraphics[width=1.0\linewidth]{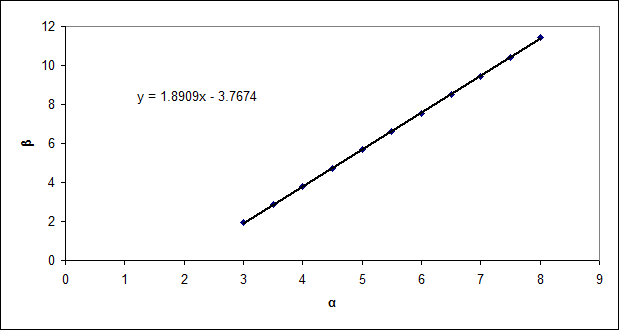}
\caption{$\beta$ vs.\ $\alpha$ with $\mu=1$ 
for the regulated potential with antisymmetric 
boundary conditions with a linear fit ($R^2=1$).}
\label{x2rBvsA}
\end{center}
\end{figure}

For $\lambda > 1/4$, we find that the eigenvalues $E_n$ 
depend on $n$ exponentially, 
except for the lowest few eigenvalues ($n=0,1$). 
The coefficient in the 
exponent is in close agreement with 
the exact Schr\"odinger spectrum (\ref{tower}) 
and with the Bohr-Sommerfeld polymer spectrum (\ref{eq:pol-WKB-evs}) 
with $\leff = \lambda - 1/4$. 
Representative spectra are shown as semi-log plots
in Figures \ref{x2urEvsn} and~\ref{x2rEvsn}, 
where the linear fit is computed 
using only the points with $n\geq2$. 
By matching the linear fit to 
the Schr\"odinger spectrum 
(\ref{tower}) and reading off the self-adjointness parameter~$\beta$, 
we can determine the self-adjoint extension of the Schr\"odinger
Hamiltonian that matches the polymer theory for the highly-excited
states. The results, shown in Figures \ref{x2urBvsA}
and~\ref{x2rBvsA}, show that the self-adjointness parameter $\beta$
depends linearly on the coupling parameter~$\alpha$,
and the slope in this relation is within 10 per cent of the 
slope obtained from the 
Bohr-Sommerfeld estimate~\nr{eq:BS-beta}, 
$\ln 8 \approx 2.0794$ (for $\mu =1$, $\mmuu =1/2$). 

Finally, our numerical eigenvalues $E_n$ are in excellent agreement
with the analytic approximation scheme of~\cite{GKS}, provided this
scheme is understood as the limit of large $\lambda$
with fixed~$n$. If the numerical results shown in figures
\ref{x2urBvsA} and \ref{x2rBvsA} are indicative of the complementary
limit of large $n$ with fixed~$\lambda$, they show that the
approximation scheme of \cite{GKS} does not extend to this limit.

\section{Conclusions}
\label{sec:conclusions}

We have compared Schr\"odinger and polymer quantizations of the
$1/r^2$ potential on the positive real line. The broad conclusion is
that these quantization schemes are in excellent agreement for the
highly-excited states and differ significantly only for the low-lying
states.  In particular, the
polymer spectrum is bounded below, whereas the Schr\"odinger spectrum
is known to be unbounded below when the coefficient of the potential
term is sufficiently negative. We also find that the Bohr-Sommerfeld
semiclassical quantization condition reproduces correctly the
distribution of the highly-excited polymer eigenvalues. At some level
this agreement is not surprising, since one expects that for any
mathematically consistent quantization scheme, in some appropriate
large~$n$, semi-classical limit, the spectra should agree. For
anti-symmetric boundary conditions both Schr\"odinger and regulated
and unregulated polymer obey the criteria, 
so it is perhaps not surprising
that they agree at least for energies close to zero. It is
somewhat surprising that they agree so well for low $n$ (where "low"
is in the context of the polymer spectra which are bounded below).

A central conceptual point was the regularization of the classical
$r=0$ singularity in the polymer theory. We did this first by
explicitly regulating the potential, using a finite differencing
scheme that mimics the Thiemann trick used with the inverse triad
operators in LQG~\cite{Thiemann}: this method allows a choice of
either symmetric or antisymmetric boundary conditions at the
origin. We then observed, as prevously noted in~\cite{GKS}, that the
singularity can alternatively be handled by leaving the potential
unchanged but just imposing the antisymmetric boundary condition at
the origin.  The numerics showed that all of these three options gave
very similar spectra, and the agreement was excellent for the
highly-excited states.

To what extent is the agreement of these three regularization options
specific to the $1/r^2$ potential?  Consider the polymer quantization
of the Coulomb potential, $-1/r$.  When the Coulomb potential is
explicitly regulated, it was shown in \cite{HLW} that the choice
between the symmetric and antisymmetric boundary conditions makes a
significant difference for the ground state energy. We have now
computed numerically the lowest five eigenenergies for the unregulated
$-1/r$ potential with the antisymmetric boundary condition, with the
results shown in Table~\ref{xurev}. Comparison with the results in
\cite{HLW} shows that the regularization of the potential makes no
significant difference with the antisymmetric boundary condition. As
noted in \cite{HLW}, for sufficiently small lattice spacing the
antisymmetric boundary condition spectrum tends to that which is
obtained in Schr\"odinger quantization with the conventional hydrogen
s-wave boundary condition~\cite{fewster-hydrogen}.

\begin{table}[htb]
\begin{center}
\begin{tabular}{|c|c|}
\hline
$n$ & $E_n$\\
\hline
0&-0.250\\
1&-0.0625\\
2&-0.0278\\
3&-0.157\\
4&-0.0100\\
\hline
\end{tabular}
\end{center}
\caption{The lowest five eigenvalues for the 
unregulated Coulomb potential with the 
antisymmetric boundary condition ($\lambda=1$, $\mu=0.1$).}
\label{xurev}
\end{table}

We conclude that in polymer quantization of certain singular
potentials, a suitably-chosen boundary condition suffices to produce a
well-defined and arguably physically acceptable quantum theory,
without the need to explicitly modify the classical potential near its
singularity: the antisymmetric boundary condition effectively removes
the $r=0$ eigenstate from the domain of the operator $1/r^2$ by
requiring $c_0=0$ in the basis state expansion $\sum_m c_m |m\mu
\rangle$. A similar observation has been made previously in polymer
quantization of a class of cosmological models, as a way to obtain
singularity avoidance without recourse to the Thiemann trick
\cite{ashtekar07,corichi08}, 
and related discussion of the self-adjointness of polymer Hamiltonians
arising in the cosmological context has been given in~\cite{Kaminski07}. 
While we are not aware of a way to relate
our system, with the $-1/r^2$ potential and no Hamiltonian
constraints, directly to a specific cosmological model, it is
nonetheless reassuring that the various techniques we have used in
this case for dealing with the singularity all lead to quantitatively
similar spectra.  Whether this continues to hold in polymer
quantization of theories that are more closely related to LQG is an
important open question that is currently under
investigation~\cite{KLZ}.

\section*{Acknowledgements}

We thank Jack Gegenberg, Sven Gnutzmann and Viqar Husain for helpful
discussions and correspondence. GK and JZ were supported in part by
the Natural Sciences and Engineering Research Council of Canada. JL
was supported in part by STFC (UK) grant PP/D507358/1.

\newpage

\begin{thebibliography}{99}


\bibitem{Thiemann07}
T.~Thiemann, 
\textit{Modern Canonical Quantum General Relativity\/}
(Cambridge University Press, Cambridge, 2007).

\bibitem{cosm sing res}
A.~Ashtekar, T.~Pawlowski and P.~Singh,
  ``Quantum nature of the big bang: Improved dynamics,''
  Phys.\ Rev.\  D {\bf 74}, 084003 (2006)
  [arXiv:gr-qc/0607039]; 
A.~Ashtekar, T.~Pawlowski, P.~Singh and K.~Vandersloot,
  ``Loop quantum cosmology of k = 1 FRW models,''
  Phys.\ Rev.\  D {\bf 75}, 024035 (2007)
  [arXiv:gr-qc/0612104]; 
K.~Vandersloot,
  ``Loop quantum cosmology and the $k = -1$ RW model,''
  Phys.\ Rev.\  D {\bf 75}, 023523 (2007)
  [arXiv:gr-qc/0612070]; 
M.~V.~Battisti, O.~M.~Lecian and G.~Montani,
  ``Polymer quantum dynamics of the Taub universe,''
Phys.\ Rev.\  D {\bf 78}, 103514 (2008) 
[arXiv:0806.0768 [gr-qc]].

\bibitem{MartinBenito:2008ej}
  M.~Martin-Benito, L.~J.~Garay and G.~A.~Mena Marugan,
  ``Hybrid Quantum Gowdy Cosmology: Combining Loop and Fock Quantizations,''
  Phys.\ Rev.\  D {\bf 78}, 083516 (2008)
  [arXiv:0804.1098 [gr-qc]].

\bibitem{husain06} 
V.~Husain and O.~Winkler,
  ``Quantum Hamiltonian for gravitational collapse,''
  Phys.\ Rev.\  D {\bf 73}, 124007 (2006)
  [arXiv:gr-qc/0601082].

\bibitem{Ashtekar:2005qt}
  A.~Ashtekar and M.~Bojowald,
  ``Quantum geometry and the Schwarzschild singularity,''
  Class.\ Quant.\ Grav.\  {\bf 23}, 391 (2006)
  [arXiv:gr-qc/0509075].

\bibitem{bh-interior}
L.~Modesto,
``Loop quantum black hole,''
Class.\ Quant.\ Grav.\  {\bf 23}, 5587 (2006)
[arXiv:gr-qc/0509078]; 
  ``Black hole interior from loop quantum gravity,''
  arXiv:gr-qc/0611043; 
C.~G.~Boehmer and K.~Vandersloot,
  ``Loop Quantum Dynamics of the Schwarzschild Interior,''
  Phys.\ Rev.\  D {\bf 76}, 104030 (2007)
  [arXiv:0709.2129 [gr-qc]]; 
  ``Stability of the Schwarzschild Interior in Loop Quantum Gravity,''
Phys.\ Rev.\  D {\bf 78}, 067501 (2008)
  [arXiv:0807.3042 [gr-qc]]; 
M.~Campiglia, R.~Gambini and J.~Pullin,
  ``Loop quantization of spherically symmetric midi-superspaces: the interior
  problem,''
  AIP Conf.\ Proc.\  {\bf 977}, 52 (2008)
  [arXiv:0712.0817 [gr-qc]]; 
R.~Gambini and J.~Pullin,
  ``Black holes in loop quantum gravity: the complete space-time,''
  Phys.\ Rev.\ Lett.\  {\bf 101}, 161301 (2008)
  [arXiv:0805.1187 [gr-qc]].

\bibitem{Thiemann}
T.~Thiemann,
  ``Quantum spin dynamics (QSD),''
  Class.\ Quant.\ Grav.\  {\bf 15}, 839 (1998)
  [arXiv:gr-qc/9606089].

\bibitem{AFW}
A.~Ashtekar, S.~Fairhurst and J.~L.~Willis,
  ``Quantum gravity, shadow states, and quantum mechanics,''
  Class.\ Quant.\ Grav.\  {\bf 20}, 1031 (2003)
  [arXiv:gr-qc/0207106].

\bibitem{halvorson} 
H.~Halvorson, 
``Complementarity of representations in quantum
mechanics,'' 
Studies Hist.\ Philos.\ Mod.\ Phys.\ {\bf 35}, 
45
(2004)
[arXiv:quant-ph/0110102]. 

\bibitem{fewster-hydrogen}
  C.~J.~Fewster,
  ``On the energy levels of the hydrogen atom,''
  arXiv:hep-th/9305102.

\bibitem{louko96}  
J.~Louko and J.~M\"akel\"a,
  ``Area spectrum of the Schwarzschild black hole,''
  Phys.\ Rev.\  D {\bf 54}, 4982 (1996)
  [arXiv:gr-qc/9605058].

\bibitem{HLW}
V.~Husain, J.~Louko and O.~Winkler,
  ``Quantum gravity and the Coulomb potential,''
  Phys.\ Rev.\  D {\bf 76}, 084002 (2007)
  [arXiv:0707.0273 [gr-qc]].

\bibitem{motl-quasi1}
  L.~Motl,
  ``An analytical computation of asymptotic Schwarzschild quasinormal
  frequencies,''
  Adv.\ Theor.\ Math.\ Phys.\  {\bf 6}, 1135 (2003)
  [arXiv:gr-qc/0212096].

\bibitem{gk-quasi1}
  G.~Kunstatter,
  ``d-dimensional black hole entropy spectrum from quasi-normal modes,''
  Phys.\ Rev.\ Lett.\  {\bf 90}, 161301 (2003)
  [arXiv:gr-qc/0212014].

\bibitem{camblong-bh}
H.~E.~Camblong and C.~R.~Ordonez,
  ``Black hole thermodynamics from near-horizon conformal quantum mechanics,''
  Phys.\ Rev.\  D {\bf 71}, 104029 (2005)
  [arXiv:hep-th/0411008].

\bibitem{GKS}
J.~Gegenberg, G.~Kunstatter and R.~D.~Small,
  ``Quantum structure of space near a black hole horizon,''
  Class.\ Quant.\ Grav.\  {\bf 23}, 6087 (2006)
  [arXiv:gr-qc/0606002].

\bibitem{case}
K.~M. Case, 
``Singular Potentials,''
Phys.\ Rev.\ {\bf 80}, 797 
(1950). 

\bibitem{frank-potential}
W.~M. Frank, D.~J. Land and R.~M. Spector, 
``Singular Potentials,''
Rev.\ Mod.\ Phys.\ {\bf 43}, 36 (1971). 

\bibitem{narnhofer} 
H.~Narnhofer, 
``Quantum theory for $1/r\sp{2}$-potentials,''
Acta Phys.\ Austriaca {\bf 40}, 306
(1974). 

\bibitem{Gupta:1993id}
  K.~S.~Gupta and S.~G.~Rajeev,
  ``Renormalization in quantum mechanics,''
  Phys.\ Rev.\  D {\bf 48}, 5940 (1993)
  [arXiv:hep-th/9305052].

\bibitem{camblong00} 
H.~E.~Camblong, L.~N.~Epele, H.~Fanchiotti and C.~A.~Garcia Canal,
  ``Renormalization of the inverse square potential,''
  Phys.\ Rev.\ Lett.\  {\bf 85}, 1590 (2000)
  [arXiv:hep-th/0003014]; 
  ``Dimensional transmutation and dimensional regularization in quantum
  mechanics. I: General theory,''
  Annals Phys.\  {\bf 287}, 14 (2001)
  [arXiv:hep-th/0003255].

\bibitem{ordonez}
H.~E.~Camblong and C.~R.~Ordonez,
  ``Renormalization in conformal quantum mechanics,''
  Phys.\ Lett.\  A {\bf 345}, 22 (2005)
  [arXiv:hep-th/0305035]; 
  ``Anomaly in conformal quantum mechanics: From molecular physics to black
  holes,''
  Phys.\ Rev.\  D {\bf 68}, 125013 (2003)
  [arXiv:hep-th/0303166]; 
G.~N.~J.~Ananos, H.~E.~Camblong and C.~R.~Ordonez,
  ``SO(2,1) conformal anomaly: Beyond contact interactions,''
  Phys.\ Rev.\  D {\bf 68}, 025006 (2003)
  [arXiv:hep-th/0302197].

\bibitem{inequivalence} 
H.~E.~Camblong, L.~N.~Epele, H.~Fanchiotti, C.~A.~Garcia Canal and C.~R.~Ordonez,
  ``On the Inequivalence of Renormalization and Self-Adjoint Extensions for
  Quantum Singular Interactions,''
  Phys.\ Lett.\  A {\bf 364}, 458 (2007)
  [arXiv:hep-th/0604018].

\bibitem{fulop07} 
\begin{flushleft}
T. F\'ul\"op, \textit{Symmetry, Integrability and Geometry, 
Proceedings of the 3rd Microconference ``Analytic and Algebraic Methods III''}, 
http://www.emis.de/journals/SIGMA/Prague2007.html 
\end{flushleft}

\bibitem{thirring-quantumbook} 
W.~E. Thirring, 
\textit{Quantum mathematical physics\/}, 2nd edition 
(Springer, Berlin, 2002). 

\bibitem{AS} 
\textit{Handbook of Mathematical Functions\/}, 
edited by M.~Abramowitz and I.~A. Stegun
(Dover, New York, 1965).

\bibitem{corichi07}
A.~Corichi, T.~Vukasinac and J.~A.~Zapata,
  ``Polymer Quantum Mechanics and its Continuum Limit,''
  Phys.\ Rev.\  D {\bf 76}, 044016 (2007)
  [arXiv:0704.0007 [gr-qc]].

\bibitem{AH}
N.~Andersson and C.~J.~Howls,
  ``The asymptotic quasinormal mode spectrum of non-rotating black holes,''
  Class.\ Quant.\ Grav.\  {\bf 21}, 1623 (2004)
  [arXiv:gr-qc/0307020].

\bibitem{gradstein} 
I.~S.
Gradshteyn
and
I.~M.
Ryzhik,
{\it Table of Integrals, Series and Products\/},
4th edition
(Academic, New York, 1980).

\bibitem{reed-simonII}
M.~Reed and B.~Simon, 
{\it Methods of Modern Mathematical Physics II: 
Fourier Analysis, Self-adjointness\/}
(Academic, New York, 1975).

\bibitem{elaydi}
S.~Elaydi, 
{\it An Introduction to Difference Equations\/}, 3rd edition 
(Springer, New York, 2005). 

\bibitem{ashtekar07} 
A.~Ashtekar, A.~Corichi and P.~Singh,
  ``On the robustness of key features of loop quantum cosmology,''
  Phys.\ Rev.\  D {\bf 77}, 024046 (2008)
  [arXiv:0710.3565 [gr-qc]].

\bibitem{corichi08} 
A.~Corichi and P.~Singh,
  ``Is loop quantization in cosmology unique?,''
  Phys.\ Rev.\  D {\bf 78}, 024034 (2008)
  [arXiv:0805.0136 [gr-qc]].

\bibitem{Kaminski07} 
W.~Kaminski and J.~Lewandowski,
  ``The flat FRW model in LQC: the self-adjointness,''
  Class.\ Quant.\ Grav.\  {\bf 25}, 035001 (2008)
  [arXiv:0709.3120 [gr-qc]].

\bibitem{KLZ}
G.~Kunstatter, J.~Louko and J.~Ziprick, 
in progress.


\end{thebibliography}
\end{document}